\documentclass[
reprint,
amsmath,amssymb,
]{revtex4-2}
\usepackage{xcolor}
\usepackage{graphicx}
\usepackage{dcolumn}
\usepackage{bm}
\usepackage[hidelinks]{hyperref}
\usepackage{siunitx} 
\usepackage[normalem]{ulem}

\begin{document}

\title{Squeezing oscillations in a multimode bosonic Josephson junction} 

\author{Tiantian Zhang}
 \email{tiantian.zhang@tuwien.ac.at}
 \author{Mira Maiw\"{o}ger} 
 \author{Filippo Borselli}
 \author{Yevhenii Kuriatnikov }
 \author{J\"{o}rg Schmiedmayer}
\author{Maximilian Pr\"{u}fer}%

\affiliation{%
Vienna Center for Quantum Science and Technology, Technische Universit\"{a}t Wien, Atominstitut, Vienna, Austria. 
}%

\date{\today} 

\begin{abstract} 

Quantum simulators built from ultracold atoms promise to study quantum phenomena in interacting many-body systems. 
However, it remains a challenge to experimentally prepare strongly correlated continuous systems such that the properties are dominated by quantum fluctuations. 
Here, we show how to enhance the quantum correlations in a {one-dimensional} multimode bosonic Josephson junction{, which is a quantum simulator of the sine-Gordon field theory}. Our approach is based on the ability to track the {non-equilibrium} dynamics of quantum properties.
After creating a bosonic Josephson junction at the stable fixed point of the classical phase space, we observe squeezing oscillations in the two conjugate variables. 
We show that the squeezing oscillation frequency can be tuned by more than one order of magnitude, and we are able to achieve a spin squeezing close to $10\,\text{dB}$ by utilising these oscillatory dynamics.  
The impact of improved spin squeezing is directly revealed by detecting enhanced spatial phase correlations between decoupled condensates.
Our work provides new ways for engineering correlations and entanglement in {the external degree of freedom of} interacting many-body systems.
\end{abstract}
\maketitle

{Understanding the role of quantum fluctuations and entanglement in interacting many-body systems is of critical importance for the development of quantum technologies, such as quantum metrology \cite{RevModPhys.90.035005,giovannetti2004quantum} and quantum simulation \cite{bloch2012quantum}. Especially the ability to prepare entanglement \cite{GUHNE20091} in quantum many-body systems is pivotal. In this context, ultracold atoms have proven to be a versatile platform \cite{RevModPhys.90.035005}.}

{Internal spin degrees of freedom (DoFs) offer a high level of control and using a single spatial mode spin squeezed states could be generated experimentally \cite{gross2010nonlinear,riedel2010atom,lucke2011twin, hamley2012spin}. Creating entangled quantum states in motional DoFs, for example in tunnel-coupled BECs in double wells (DWs) \cite{esteve2008squeezing,berrada_integrated_2013} is, on the contrary, less explored; {the impact of spin squeezing and entanglement has only been studied with respect to the global observables.}}

{For spatially extended systems, where more than one longitudinal mode is occupied, the interplay between {transverse (spin)} and {longitudinal}  DoFs leads to new physical phenomena. 
For coupling internal states interesting dynamical phenomena as well as a high degree of control have been shown \cite{nicklas2015Scaling,farolfi2021elBJJ}; for tunnel-coupled systems, the influence of the multimode situation on Josephson oscillations has been studied \cite{pigneur2018relaxation}. However, so far the quantum regime has not been accessible.}

\begin{figure}[h!]
    \centering
    \includegraphics[width = 0.4\textwidth]{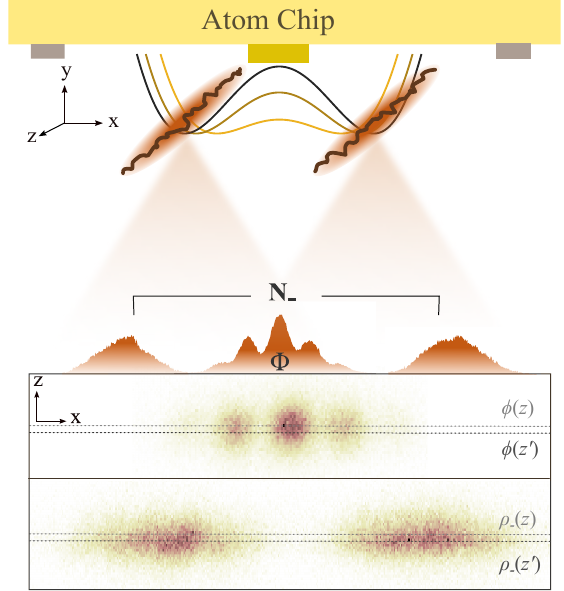}
    \caption{\textbf{System and readout with spatial resolution} Two 1D quasi-BECs {with locally fluctuating quantum phases} trapped magnetically below an Atom Chip with RF dressing technique. {The local relative phase $\phi(z)$ and relative atom density $\rho_-(z)$ between the two BECs are detected with fluorescence imaging after a long Time-of-Flight. The spatial resolution allows us to probe the spatial phase correlations along the condensates. We integrate over the longitudinal direction to obtain the global observables, namely the relative phase $\Phi$ and relative atom number $N_-$.}}
    \label{fig:expsetup}
\end{figure}

{Interestingly, in the one-dimensional (1D) regime these systems are excellent quantum simulators for the sine-Gordon field theory \cite{gritsev2007linear,coleman1975quantum}, which was mainly explored in tunnel-coupled systems \cite{schweigler2017experimental,schweigler2021decay}. It is worth pointing out that BECs in DWs at equilibrium can be experimentally achieved by direct cooling in DWs \cite{schweigler2017experimental}. But {fluctuations introduced by the finite temperature}
make it challenging to prepare the system in the quantum correlated regime \cite{paraoanu2001josephson}. However, it has been shown that using a splitting routine from a single condensate, quantum correlated states can be prepared in a multimode scenario \cite{berrada_integrated_2013}. In this work, we combine the ability to prepare quantum correlated states with spatially resolved measurements of the sine-Gordon fields entering a new regime.}

{We realise a multimode bosonic Josephson junction (BJJ) \cite{smerzi1997quantum,raghavan1999coherent} by splitting 1D quasi-BECs into a DW (see Fig.\,\ref{fig:expsetup} upper panel). Instead of studying Josephson oscillations \cite{albiez2005direct,gati2007bosonic,levy2007ac, leblanc2011dynamics,PhysRevLett.118.230403,pigneur2018relaxation}, we prepare the system at the stable fixed point of the classical phase space and access its quantum properties. Despite the stationary expectation values of both relevant observables, relative number, and relative phase, we observe oscillatory dynamics of the quantum fluctuations {which result from the evolution of an initial quantum state different from the ground state at zero temperature}. We employ these non-equilibrium dynamics of the fluctuations in the BJJ to foster strong spin squeezing in decoupled condensates.} {Furthermore, we study its impact on spatial phase correlations in the multimode system.}

{To probe the multimode properties of the prepared states we use a spatially resolved detection of the relative phase (see Fig.\,\ref{fig:expsetup} lower panel). The observations in this observable, enable us to demonstrate how the improved number squeezing enhances spatial correlations between two decoupled 1D condensates. We are able to connect quantum properties resulting from the preparation process to enhanced phase correlations in a regime where a sine-Gordon model is realized. The enhanced squeezing can in principle be directly related to lower effective temperatures in the prethermalised regime \cite{kitagawa2011dynamics,Gring_2012} and our work thus shows a pathway to prepare sine-Gordon field simulators in a regime dominated by quantum fluctuations.}

\section*{Realisation and readout of a multimode BJJ} 

\begin{figure*}[t!]
    \centering
    \includegraphics[width=0.7\textwidth]{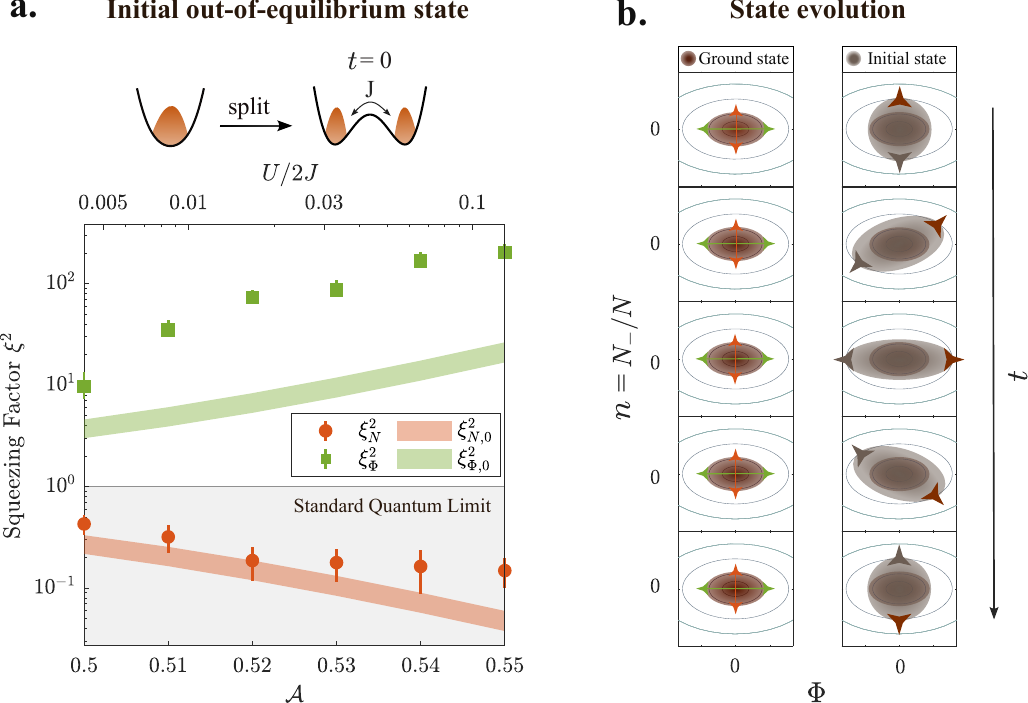}
    \caption{\textbf{Preparation and dynamics of spin-squeezed quasi-BECs in the Josephson regime}
    \textbf{a.} {Measured squeezing factors in $N_-$ and $\Phi$ quadrature, denoted as $\xi_N^2$ and $\xi_{\Phi}^2$, right after a linear ramp-up at a constant ramp speed to various tunnel-coupled DWs, denoted with the rescaled RF dressing amplitude $\mathbf{\mathcal{A}}$.} 
    In the upper x-axis, the parameter {$1/N \ll U/2J \ll N $} indicates that we are in the Josephson regime of BJJ. The coloured bands mark the expected ground state squeezing factors of BJJ at zero temperature (Eq.\,\eqref{eq:bhhc}) with atom number $N\in[2,5]\times 10^3$.
    \textbf{b.} Schematics of state evolution in the classical phase space. {Left column illustrates the expected ground state fluctuation of BJJ according to Eq.\,\eqref{eq:sgsqfac}, where orange (green) arrows represent projection noise in the relative number (phase) quadrature. The right column represents the out-of-equilibrium quantum state evolution. The triangle markers indicate single classical realizations orbiting along the equipotential lines at the plasma frequency $f_p$. In the Josephson regime, the distribution of out-of-equilibrium states rotates and deforms which results in squeezing oscillations.} A full period of the quantum state evolution around the stable fixed point ($n = N_-/N =0$ and $\Phi = 0$) conforms to a half period of evolution of single realisations (mean values) in the phase space.}
    \label{fig:squeezingplusdyn}
\end{figure*}

We realise a multimode BJJ with $1$D {tunnel-coupled quasi-}BECs of $^{87}\text{Rb}$ atoms trapped magnetically under an Atom Chip \cite{reichel2011atom}. 
The DW potential is generated with the Radio-Frequency (RF) dressing \cite{hofferberth2006radiofrequency} technique. 
{The dressing is calibrated such that} the precise transformation from a single well to a DW is achieved by ramping up the amplitude $\mathbf{\mathcal{A}}$ of the RF field whose value is normalised to the maximal current.
We display in Fig.\,\ref{fig:expsetup} the transverse trap configuration and typical experimental readout with our single-atom-sensitive fluorescence imaging system \cite{Buecker_2009} after Time-of-Flight of $t_F = 43.4\,\text{ms}$. The aspect ratio between the radial and axial trap frequency is $\sim 100$. 

{Each of the quasi-condensate in a DW can be described in the density-phase representation by $\Psi_j = \sqrt{\rho_j(z)}\times\text{exp}(-i\phi_j(z))$, where $j \in [L,R]$ labels the left ($L$) and the right ($R$) condensate. The local density $\rho_j$ and local phase $\phi_j$ are spatially varying and fluctuating fields. This is illustrated by the wiggly brown lines in Fig.\,\ref{fig:expsetup}. We are interested in the local fields in the relative DoF, namely the local relative density $\rho_{-}(z) = \rho_{R}(z) - \rho_{L}(z)$ and local relative phase $\phi(z) = \phi_{R}(z) - \phi_{L}(z)$. For finitely tunnel-coupled condensates, the spatially dependent relative phase field $\phi(z)$ can be described by the sine-Gordon field theory \cite{gritsev2007linear}.}\par

{Our imaging allows us to measure the relative phase with spatial resolution (see Fig.\,\ref{fig:expsetup} lower panel). From this, we can access the two-point phase correlations to study the sine-Gordon physics in the multimode regime, as will be discussed later. Valuable insights into the dynamical quantum properties of BJJs can already be obtained from a single-mode description of the BJJ \cite{PhysRevA.55.4330}. Therefore, we will first investigate the properties of the two global observables of the 1D BJJ, the relative atom number, $N_- = \sum_z \rho_-(z)$ and the relative phase, $ \Phi = \sum_z \phi(z)$, which correspond to the spatial zeroth mode.}

The macroscopic dynamics {of the global observables $N_-$ and $\Phi$} in a BJJ can be described by the two-mode Bose-Hubbard (BH) model
\begin{equation}\label{eq:bose-hubbard}  
 \mathcal{H} = \frac{2J}{\hbar} \left[ \frac{UN}{4J}  n^2 - \sqrt{1- n^2} \cos {\Phi}\right],
\end{equation} 
where $N = N_\text{L}+N_\text{R}$ is the total atom number, $ n= N_-/ N$ is the relative imbalance, $U$ is the interaction strength and $J$ is the single particle tunnel coupling strength. {The Josephson regime of the BJJ is indicated by {$1/N \ll U/2J \ll N$}.}  
From Eq.\,\eqref{eq:bose-hubbard}, we can obtain the Josephson oscillation frequency \cite{smerzi1997quantum,raghavan1999coherent} 
of the mean values $\langle \Phi \rangle$ and $\langle N_- \rangle$  
\begin{equation} \label{eq:plasma}
    f_p = \frac{2J}{h} \sqrt{\langle \cos \Phi_0 \rangle+\frac{UN}{2J}},
\end{equation}
also known as the plasma frequency. Here $\cos \Phi_0$ indicates the initial phase coherence factor. 
From Eq.\eqref{eq:plasma}, we see that $f_p$ depends explicitly on the single particle tunnel coupling strength $J$.\par 

{For the multimode many-body dynamics dominating in decoupled DWs, the two-mode BH model does definitely not give a full description. However, we will see in the following that it nicely captures the dynamics of the squeezing in the spatial zeroth mode. Exact modelling of the splitting process together with the ensuing dynamics are challenging when taking into account the true multimode quantum dynamics; thus, quantum simulations can provide new insights.}\par

\section*{Preparing quantum-correlated states}
To prepare strongly correlated BECs, we split a single BEC into two {by transforming a single well into a DW (see Fig.\,\ref{fig:squeezingplusdyn}\textbf{a})}. BEC splitting as a pathway to generate quantum correlated states was investigated in many earlier works \cite{maussang2010enhanced,esteve2008squeezing,jo2007long}; {here, we introduce and implement a new scheme which is based on understanding the non-equilibrium dynamics after a non-adiabatic splitting into a tunnel-coupled DW}. In the following, we introduce the relevant quantities and summarize our findings which lay the basis for observing the non-equilibrium dynamics of the prepared squeezed state.

To quantify the fluctuations of the observables, we define the squeezing factors 
\begin{equation} \label{eq:squeezing}
   \xi_N^2  = \frac{\Delta^2 N_-}{N} , \hspace{5pt}
   \xi_\Phi^2 = \Delta^2 \Phi\cdot N , \\
\end{equation} 
where $\Delta^2 N_-$ and $\Delta^2 \Phi$ represent the statistical variance{, evaluated as in Ext.\,Data Fig.\,\ref{fig:histos}.} We use the quadrature projection noise of spin coherent states in the denominators. 
Hence $\xi_N^2 = \xi^2_\Phi = 1$ represents the standard quantum limit. Furthermore, spin-squeezed states \cite{wineland1994squeezed,sorensen2001many} representing a class of entangled states are characterised by {spin squeezing factor} $\xi_s^2 = \xi^2_N/\langle \cos \Phi \rangle^2  < 1$, where $\langle \cos \Phi \rangle$ is the phase coherence factor.\par

The ground state of the many-body Hamiltonian \cite{milburn1997quantum} exhibits a growing degree of number squeezing in less coupled DWs owing to repulsive interatomic interactions. 
In Fig.\,\ref{fig:squeezingplusdyn}\textbf{a}, we show the experimentally inferred squeezing factors in $N_-$ and $\Phi$ as a result of BEC splitting with finite duration, denoted with rescaled RF dressing amplitude $\mathbf{\mathcal{A}}$. 
We conduct $\sim 200$ repetitions for each measurement to ensure reliable statistics. {At the beginning of the Josephson regime, $U/2J \sim 1/N$, the measured relative number fluctuations of split BECs approach the expected ground state number squeezing $\xi^2_{N,0}$ (see Appendix Eq.\,\eqref{eq:sgsqfac}).}\par

During further ramp-up of the RF amplitude $\mathbf{\mathcal{A}}$, the adiabaticity condition can no longer be satisfied. This breakdown is shown explicitly in Fig.\,\ref{fig:squeezingplusdyn}\textbf{a}, {where $\xi_N^2$ become increasingly larger than $\xi_{N,0}^2$ in less tunnel-coupled DWs}. This initialised out-of-equilibrium state with BEC splitting is expected to evolve dynamically in the Josephson regime (see Fig.\,\ref{fig:squeezingplusdyn}\textbf{b}). {As we will discuss in the next sections, this evolution of the quantum state in phase spaces with number-squeezed ground states is not just a rotation, but also additional shearing.}\par

The fluctuations in the relative phase are always much above the expected ground state phase squeezing factor $\xi_{\Phi,0}^2$ as shown in Fig.\,\ref{fig:squeezingplusdyn}\textbf{b}. This is due to the interatomic interaction induced phase diffusion \cite{lewenstein1996quantum} during the ramp and can be prevented by increasing the splitting speed {of linear ramps}. The trade-off of faster splitting is poorer number squeezing right after the splitting. {Thus alternative splitting routines beyond linear single ramps are sought after to enhance spin squeezing in decoupled traps.}


\section*{Squeezing oscillations in conjugate quadratures}
\begin{figure}[h!]
    \centering
    \includegraphics[width=0.4\textwidth]{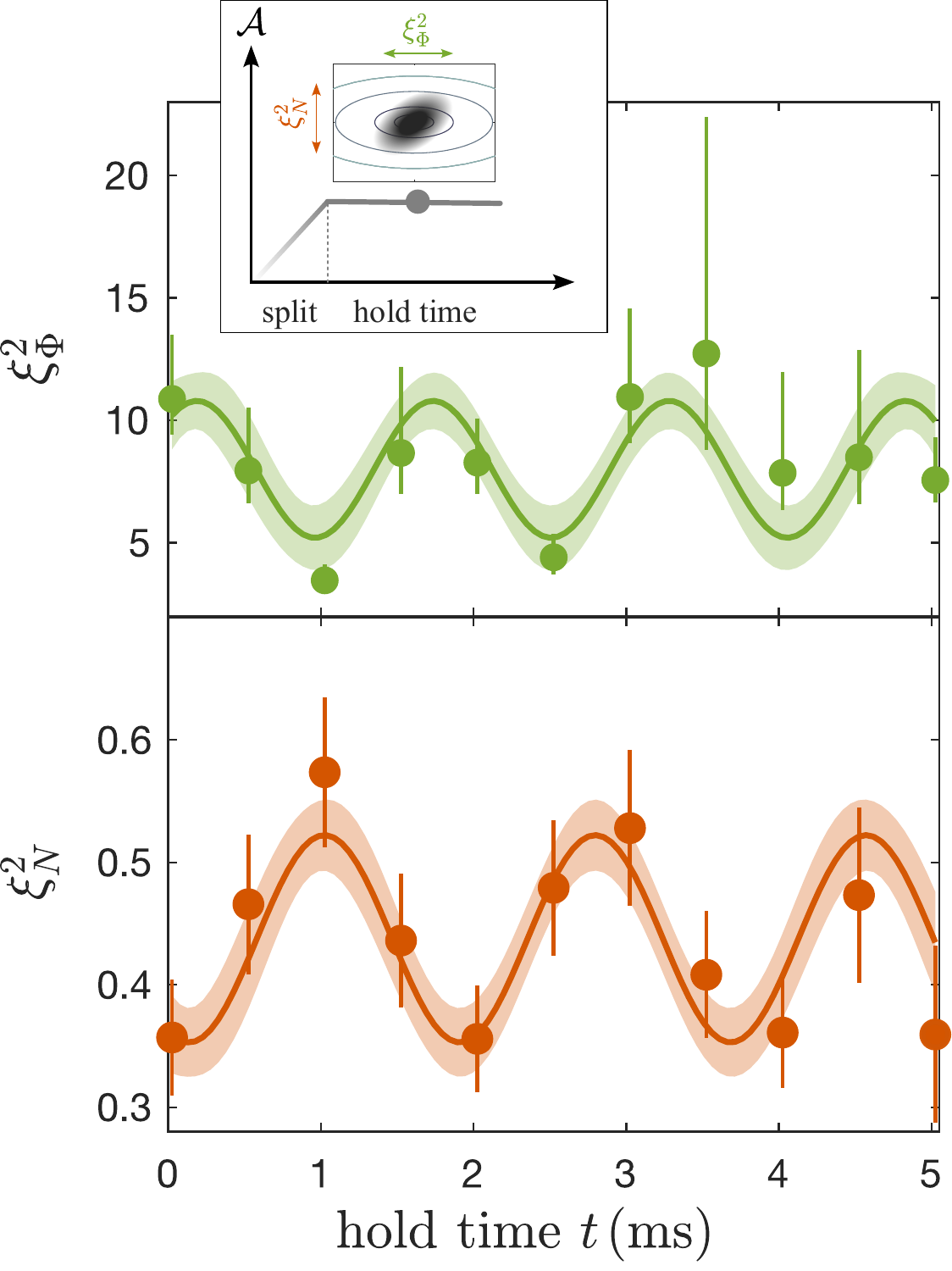}
    \caption{\textbf{Squeezing oscillations in a coupled double well}. 
    \textbf{a.} Dynamics of the fluctuations of relative phase (green) and relative number (orange markers) for variable hold time in the coupled trap after the splitting. We observe oscillatory dynamics with comparable frequency and relative phase shift of $\pi$. {The oscillatory dynamics is not a simple linear rotation of the state in phase space but results from the interplay of tunnel-coupling and onsite interaction and leads to a rotation plus deformation of the state (see inset).} From a sinusoidal fit (solid lines) we extract the frequencies which are {roughly} twice the plasma frequency as expected from the rotation and shearing of the distribution in the phase space (see inset and Fig\,\ref{fig:squeezingplusdyn}\textbf{b}). Bands indicate $68\%$ prediction confidence interval of the fits and errorbars represent one s.\,e.\,m.}
    \label{fig:conjosci}
\end{figure}

We prepare two BECs in a strongly coupled DW {($\mathbf{\mathcal{A}} = 0.5$)} by linearly ramping up from a single well; this prepares the system at the stable fixed point of the classical phase space, i.\,e. $\langle N_-\rangle = 0$ and $\langle \Phi\rangle = 0$ {(see Ext.\,Data Fig.\,\ref{fig:JosephsonExample})} with phase space fluctuations different from the ground state. 
By tracking the evolution of this out-of-equilibrium quantum state in the strongly coupled double well, we observe the dynamics of the quantum fluctuations of the conjugate observables (see Fig.\,\ref{fig:conjosci}). 

{The squeezing factors in both quadratures undergo oscillatory dynamics. Strikingly, the number quadrature stays always squeezed while the phase never gets squeezed below the SQL, i.e. the oscillatory dynamics is not a simple linear rotation of the state in phase space. This results from the interplay of tunnel coupling and onsite interaction and leads to a rotation and deformation of the state (see Fig.\,\ref{fig:squeezingplusdyn}\textbf{b} and Fig.\,\ref{fig:conjosci} inset). We can understand the frequency of the oscillation with the intuitive picture that a $\pi$ rotation of a single realisation corresponds to a $2\pi$ rotation of quantum state distribution in phase space (see Fig.\,\ref{fig:squeezingplusdyn}\textbf{b} and Appendix). We thereby deduce that the squeezing oscillations are twice as fast as the Josephson oscillations of the mean (with plasma frequency $f_p$); this is in accordance with a semiclassical analysis for Raman coupled BECs \cite{dunningham2001relative}.}
\par 

We fit the observed squeezing factors in Fig.\,\ref{fig:conjosci} with a sine function to determine the squeezing oscillation frequency and obtain $f_\xi = 567(29)\,\text{Hz}$ in relative number $N_-$ quadrature with total atom number $N = 4154(35)$ and $f_\xi = 649(33)\,\text{Hz}$ in relative phase $\Phi$ quadrature with $N = 4302(45)$ (see Appendix).  
The measured squeezing oscillations in both quadratures as expected from the nonlinear Josephson dynamics match with twice the experimentally measured plasma frequency $f_p$ and have a $\pi$ phase shift with respect to each other.  
Combining the complementary measurements in Fig.\,\ref{fig:conjosci}, we infer the mean quantum state fluctuations in the phase space to be $\overline{\xi_N^2}\cdot \overline{\xi_\Phi^2} \approx 3.5$ with $\overline{\xi_N^2} = 0.44(2)$ and $\overline{\xi_\Phi^2} = 8.0(7)$.\par

\section*{Two-step sequence for optimised spin squeezing}

\begin{figure}[t!]
    \centering
    \includegraphics[width = 0.5\textwidth]{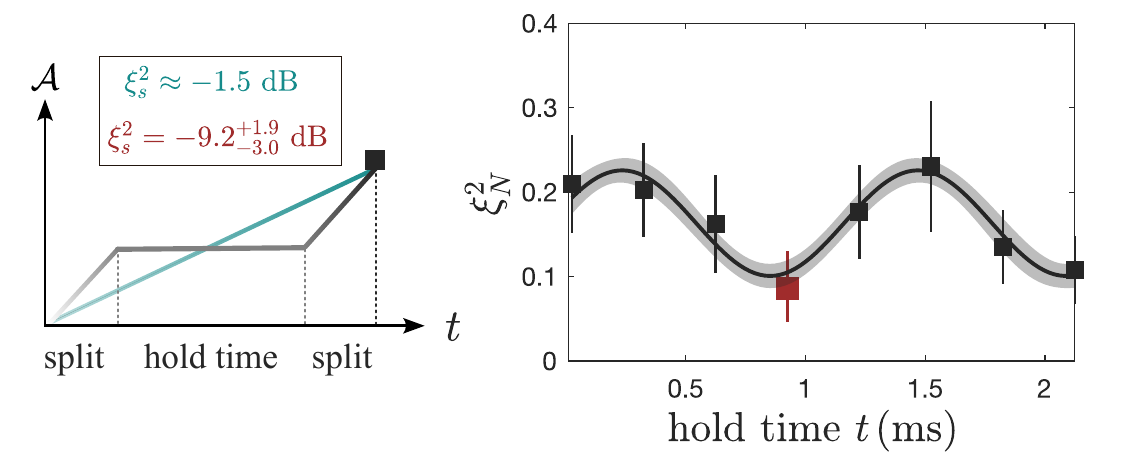}
    \caption{\textbf{Two-step sequence for optimised spin squeezing} For optimising the spin squeezing in the decoupled trap ($\mathbf{\mathcal{A}} = 0.65$), we perform a two-step sequence (black line in left schematic) instead of a simple linear ramp (blue line). {By varying the hold time in the coupled trap ($\mathbf{\mathcal{A}} = 0.5$) and performing the final measurement after adding a second linear ramp to the decoupled trap, we observe an oscillatory behaviour. This oscillation is similar to Fig.\,\ref{fig:conjosci}, but overall better number squeezing.} For the optimal spin squeezed point (red marker) we find $-9.2\,$dB spin squeezing instead of $-1.5$\,dB for a single linear ramp {with a similar total duration as the two-step sequence}.}
    \label{fig:two-step}
\end{figure}

So far mostly simple linear ramps have been used to generate squeezing. In single-mode situations theoretical studies showed that non-linear ramps generated by optimal control can lead to better squeezing \cite{PhysRevA.92.053632,PhysRevA.80.053625}; for our multimode BJJ the development of a many-body simulation, required for performing open loop optimisation {on spin squeezing}, is not possible, as the splitting process is hard to calculate. 
Thus we develop an experimentally tractable two-step approach (see Fig.\,\ref{fig:two-step}) based on the observed squeezing oscillations in the coupled DW; in single-mode spinor BECs a similar approach has been used to achieve spin-squeezed ground states \cite{PRXQuantum.3.010328,arxiv.2202.12338}. Here, we employ it for optimising the spin squeezing in a decoupled DW {($J = 0$)} in a multimode situation (see Fig.\,\ref{fig:two-step}).
This is of particular interest for their application as a sensitivity-enhanced matter-wave interferometer \cite{berrada_integrated_2013,RevModPhys.90.035005,giovannetti2004quantum}.\par 

We show in Fig.\,\ref{fig:two-step} that the generated squeezing oscillation in the coupled DW is successfully preserved after the ramp to the decoupled DW with two-step splitting; {this means we are able to manipulate the quantum properties in the decoupled trap by a holding time in the coupled trap. Furthermore, the} number squeezing is even further enhanced during the second ramp.
With this approach, we obtain a phase coherence factor of $\langle \cos \Phi \rangle = 0.86^{+0.01}_{-0.02}$.
This leads to a spin squeezing factor \cite{gross2010nonlinear} of $\xi_s^2 = -9.2^{+1.9}_{-3.0} \, \text{dB}$ (with no detection noise correction, $\xi_s^2 = -4.0(1.1) \, \text{dB}$, see Appendix), which witnesses many-body entanglement \cite{sorensen2001many}.\par 

In contrast, a single linear ramp with {a similar} duration yields lower spin squeezing of $\xi_s^2 \approx -1.5\,\text{dB}$, demonstrating a significant gain with two-step splitting.
A simple way to understand this is that two-step splitting enables us to achieve optimal number squeezing at an earlier stage of the splitting procedure and, as a result, suppressing the phase diffusion more efficiently.\par

\section*{Tuning squeezing oscillations}
To achieve controlled preparation of strongly correlated states with two-step splitting, it is crucial to understand the control parameters and tunability of the squeezing oscillation frequency. 
We identify two strategies: the first one involves tuning the parameters in a static BJJ, and the second one involves introducing time dependence on the control parameters through transversal motions.\par

\begin{figure}[t!]
    \centering
    \includegraphics[width=0.5\textwidth]{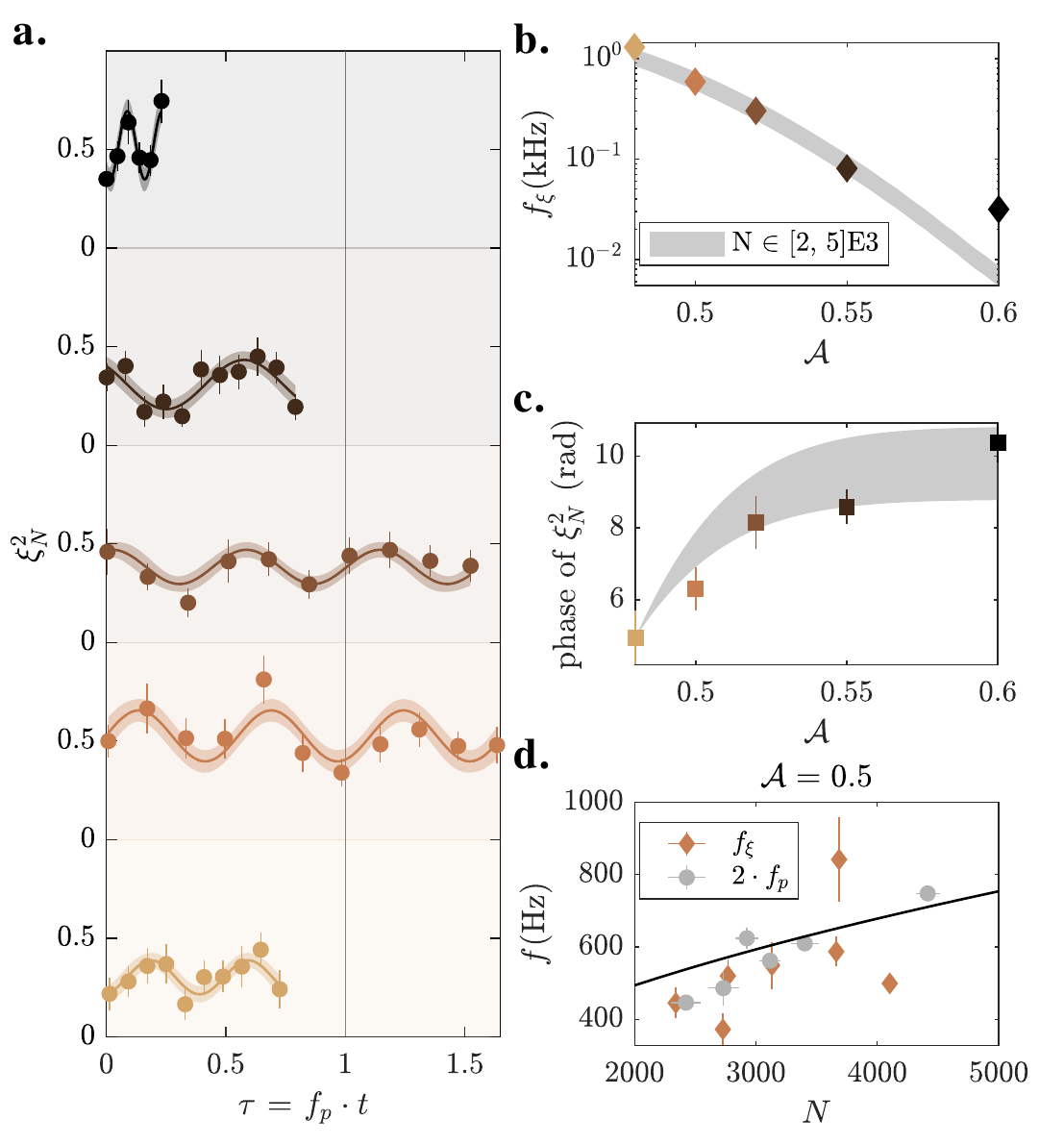}
    \caption{\textbf{Tuning squeezing oscillation frequency} \textbf{a.} Observed number squeezing factor $\xi_N^2$ evolution in coupled DWs {after a constant splitting speed} with increasing $\mathbf{\mathcal{A}}$ (bottom to top). The solid line is a fitted sine function with $68\%$ simultaneous prediction bounds. 
    \textbf{b.} The extracted squeezing frequency $f_\xi$ (diamond) from \textbf{a.} together with the calculated prediction $2f_p$ (grey shade) with $N\in [2,5]\cdot\num{e3}$. 
    {\textbf{c.} Fitted initial phases of squeezing oscillations in \textbf{a.} compared to the expected accumulated phases (grey band) of the quantum state during the linear ramp with constant ramp speed to different DWs at $\mathbf{\mathcal{A}}$ with atom number $N\in [2,5]\cdot\num{e3}$ as in \textbf{b}. Here the band uses the fitted phase of squeezing oscillation at $\mathbf{\mathcal{A}} = 0.48$ (in \textbf{a}) as the initial value.}
    \textbf{d.} Dependence of $f_\xi$ (diamond) on total atom number $N$ in coupled DW {at} $\mathbf{\mathcal{A}} = 0.5$, {and for comparison} experimentally measured plasma frequency $2 f_p$ (circle) and solid line marks the inferred $2 f_p$ from Eq.\,\eqref{eq:plasma}.}
    \label{fig:freqscaling}
\end{figure}

First, we explore experimentally the frequency scaling in elongated BJJ based on Eq.\eqref{eq:plasma}. 
In Fig.\,\ref{fig:freqscaling}\textbf{a}, we show the measured number squeezing oscillations in different DWs with {decreasing single particle tunnel coupling $J$ (bottom to top)}. 
We observe {squeezing oscillations with} frequencies spanning more than one order of magnitude. 
In Fig.\,\ref{fig:freqscaling}\textbf{b}, we plot collectively the extracted frequencies and as a comparison the expectation of $2 f_p$ (grey band) estimated directly from Eq.\,\eqref{eq:plasma}. 
Here, $J$ for each $\mathbf{\mathcal{A}}$ is inferred from simulated DW potential and experimental atom number $N \in [2, 5]\cdot\num{e3}$. 
We find good agreement between the {experimentally observed fluctuation dynamics on the global observable $N_-$ (zeroth mode) of the multimode BJJ and the simple calculations based on the two-mode BH model}. 
This confirms first of all that the observed squeezing oscillations indeed originate from stationary nonlinear Josephson dynamics.\par

{On top of which, we show in Fig.\,\ref{fig:freqscaling}\textbf{c} that fitted phases of the squeezing oscillations in Fig.\,\ref{fig:freqscaling}\textbf{a} match quantitatively with the expected evolution of the quantum state during the linear ramp. The experimental results in Fig.\,\ref{fig:freqscaling}\textbf{a} are obtained after single linear ramps with the same splitting speed. We can estimate the additionally gained phase during the ramp compared to the first experimental data point at $\mathbf{\mathcal{A}} = 0.48$ (set as $t = 0$) as the sum of constantly evolving plasma frequencies $f_p(t_i)$ (Eq.\,\eqref{eq:plasma}) over a ramp time $\Delta T$, namely $\Delta T \cdot \sum_{t=0}^{\Delta T} 2\pi f_p(t)$. Despite the quantum state evolution in the phase space (Fig.\,\ref{fig:squeezingplusdyn}\textbf{b}) not being a simple rotation, the differences between the initial phases in different DWs built up during the ramp can be surprisingly well predicted with this simple estimation.}\par

The other DoF for tuning squeezing oscillation is the total atom number $N$. 
In Fig.\,\ref{fig:freqscaling}\textbf{c} we show experimentally measured squeezing oscillation frequency $f_\xi$ as a function of $N$ (for fixed $\mathbf{\mathcal{A}} = 0.5$) and compare them directly with experimentally measured plasma frequency $2f_p$ and a solid line derived from Eq.\,\eqref{eq:plasma}. 
In this strongly coupled DW, we can tune squeezing oscillation frequency from $300\,\text{Hz} $ to $800\,\text{Hz}$ by only adjusting the atom number.\par

\begin{figure}[t!]
    \centering
    \includegraphics[width=0.5\textwidth]{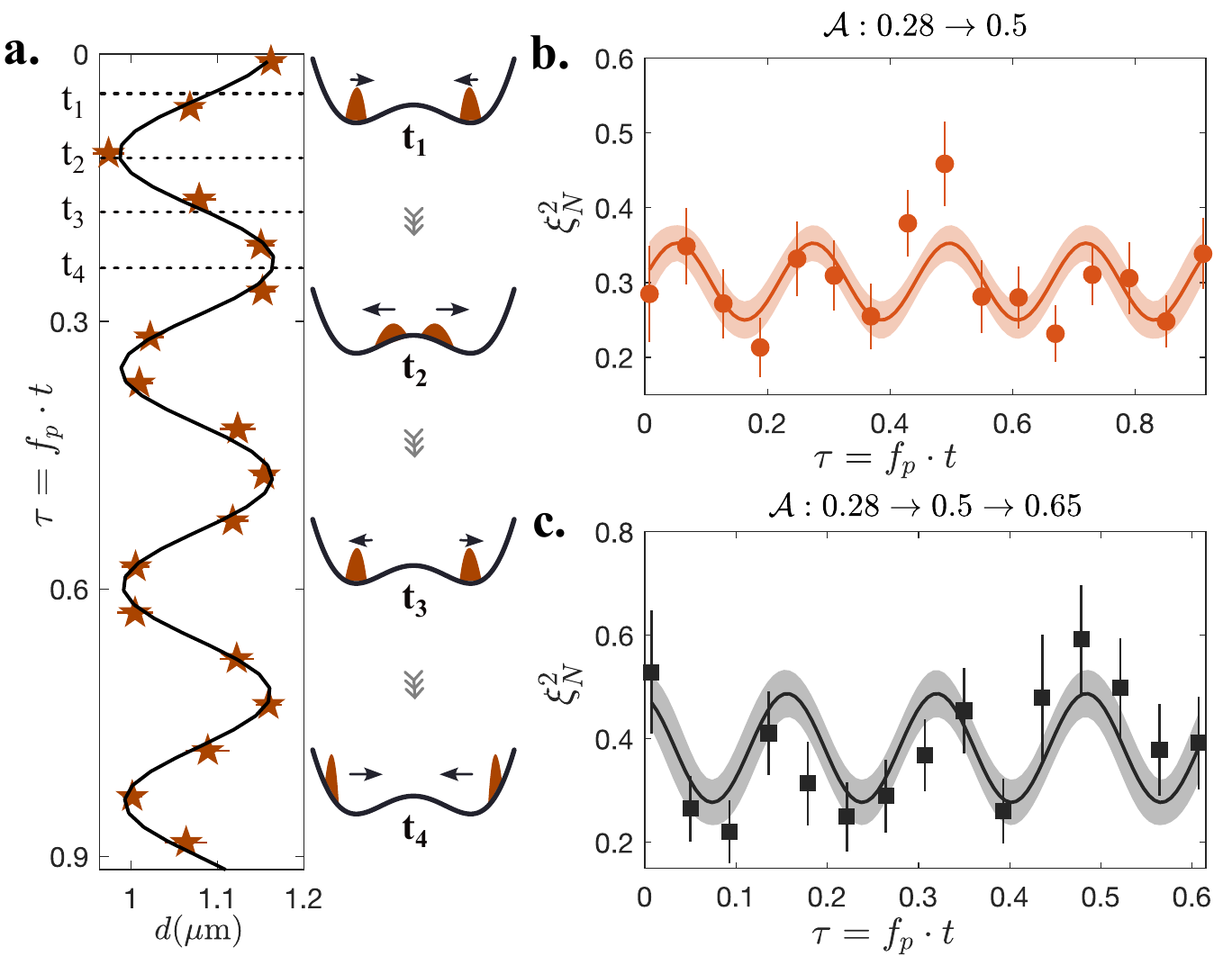}
    \caption{\textbf{Driving number squeezing oscillation}. \textbf{a.} Inter-condensate distance $d$ in $0.5$ trap resulting from splitting quench with $\kappa = 0.085\, \text{ms}^{-1}$ into $\mathbf{\mathcal{A}}=0.5$ trap. $d$ oscillates at the transverse trap frequency $f_x$. With this effective periodic modulation of tunnel coupling, we observe in \textbf{b.} that $\xi_N^2$ (orange circle) is driven at the trap frequency $f_x > f_p$. \textbf{c.} Transferred squeezing oscillation after a two-step quench. $\tau$ indicates hold time in coupled DW.} 
    \label{fig:drivefreq}
\end{figure}

To expand the tuning capabilities, we incorporate an active modulation on the tunneling mechanism.
By performing a splitting quench, we excite out-of-phase transverse sloshing between the two BECs at the transverse trap frequency $f_x = 1418(10) \, \text{Hz}$. 
We depict in Fig.\,\ref{fig:drivefreq}\textbf{a} this induced motion with the inferred inter-condensate distance $d$. 
This motional excitation drives the effective tunnel coupling $J$ periodically at the trap frequency $f_x$. 
We show in Fig.\,\ref{fig:drivefreq}\textbf{b} how this periodic drive enforces $\xi_N^2$ to oscillate at frequencies comparable with $f_x$. Furthermore, we can reproduce this oscillation frequency with a two-step splitting quench to the decoupled DW (see Fig.\,\ref{fig:drivefreq}\textbf{c}).\par%

By utilizing this method, we boost the preparation of spin-squeezed states with two-step from two perspectives: reduced ramp times and faster squeezing dynamics. 
Our observations potentially facilitate investigations into captivating phenomena like parametric resonance \cite{Parametric} and Floquet engineering \cite{ji2022floquet}.\par

\section*{Squeezing-protected spatial correlations}
{With the gained insight on how to optimise spin squeezing, we now investigate how the spin squeezing influences the spatial correlations in our multimode system which is a quantum simulator for the sine-Gordon field theory.}
{In a decoupled DW,} number squeezing prolongs the {global} phase coherence by reducing the relative phase diffusion {\cite{berrada_integrated_2013,jo2007long}}.
Indeed,  we find a good qualitative agreement between experimentally extracted global phase diffusion rates and the levels of global number squeezing (see Ext.\,Data Fig.\,\ref{fig:split_speed}). 

The spatial resolution of our imaging system grants direct access to the local relative phase (see Appendix). 
This allows us to explore how the observed number squeezing impacts the local dephasing \cite{kitagawa2011dynamics,bistritzer2007intrinsic}.
 \begin{figure}[h!]
    \centering
    \includegraphics[width=0.5\textwidth]{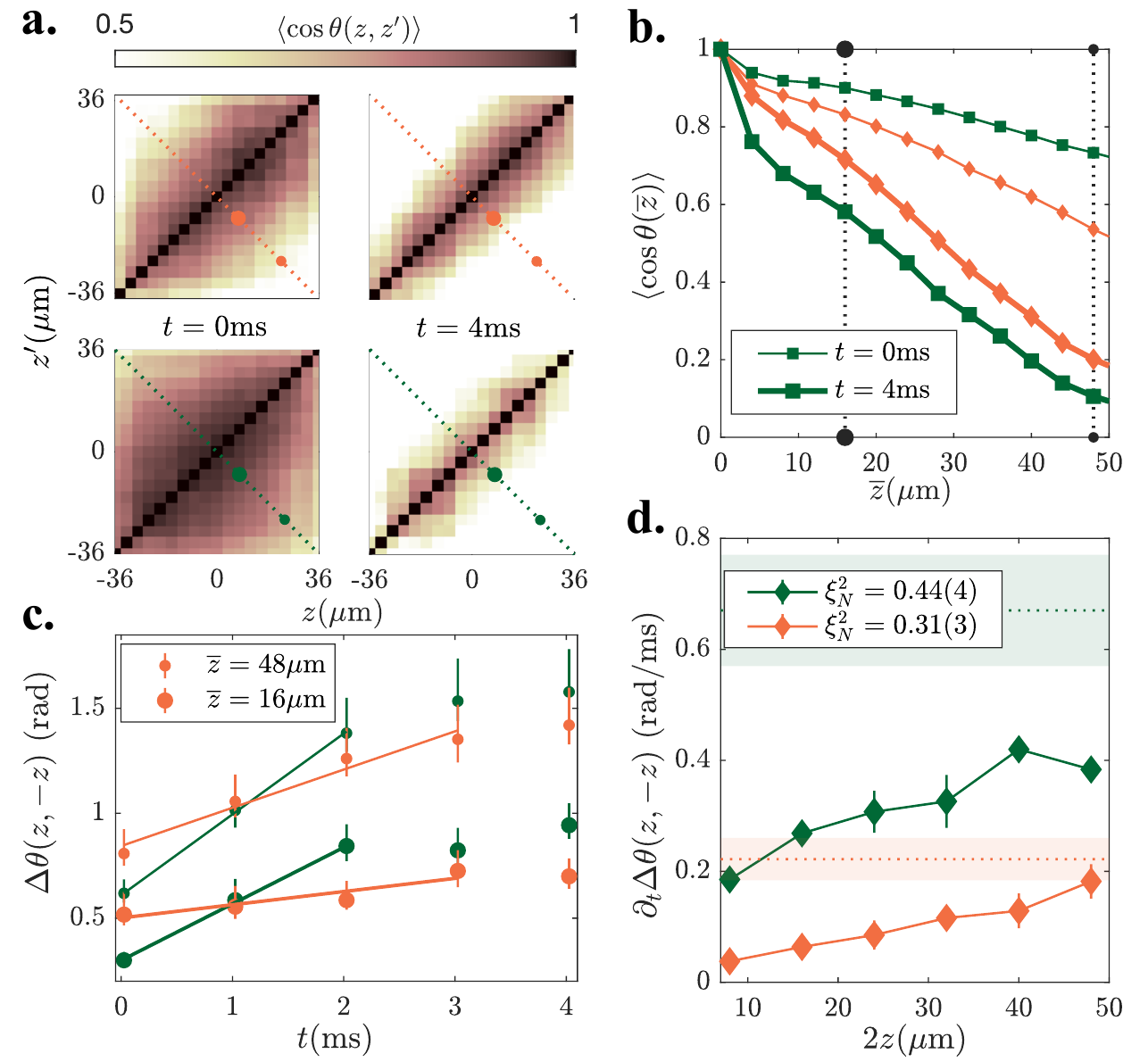} 
    \caption{\textbf{Influence of number squeezing on multimode phase correlation function.} 
    \textbf{a.} Phase correlation function PCF $\langle \cos \theta(z, z^\prime)\rangle$ between two spatial positions $z$ and $z^\prime$ along the condensates. 
    Upper and lower panels show PCF of a state with $0.31(3)$ (orange) and $\xi_N^2 = 0.44(4)$ (green) and at time $t$ after splitting to DW at $\mathbf{\mathcal{A}} = 0.6$. 
    \textbf{b.} Spatially averaged PCF, $\langle \cos \theta(\overline z) \rangle$, with $\overline z = |z - z^\prime|$ from \textbf{a.} visualises how number squeezing suppresses decay of PCF. 
    \textbf{c.} Evolution of $\Delta \theta(z,-z)$ for $z =  8 \mu \text{m}$ (big circle) and $z = 24\, \mu\text{m}$ (small circle). 
    We infer the dephasing rate $\partial_t \Delta \theta(z,-z)$ with a linear fit and plot the extracted rates in \textbf{d} (see Ext.\,Data Fig.\,\ref{fig:localphasediff} for all distances $z$). 
    The shaded regions represent the extracted global phase diffusion rate. 
    We observe a spatial dependence of $\partial_t \Delta \theta(z,-z)$ along the condensates, which hints at local number squeezing originating from the multimode dynamics of quasicondensates.}
    \label{fig:pcf}
\end{figure} 
To study this, we prepare split BECs in an effectively decoupled DW, with two different levels of (global) number squeezing. 
We consider the two-point phase correlation function (PCF), defined as $\langle \cos\theta(z, z^\prime)\rangle$, where $\theta(z, z^\prime) = \phi(z)-\phi(z^\prime)$ is the relative phase field.\par 

In Fig.\,\ref{fig:pcf}\textbf{a}, we compare the two-point PCF in the decoupled DW at two time instances, $t = 0\, \text{ms}$ and $t = 4\, \text{ms}$, after initial preparation with $\xi_N^2 = 0.31(3)$ (upper panels) and $0.44(4)$ (lower panels). Despite higher PCF in the beginning, the decay is faster with weaker number squeezing (lower panels).
For better visualisation, we plot in Fig.\,\ref{fig:pcf}\textbf{b} the averaged PCF, $\langle \cos\theta(\overline z)\rangle$, where $\overline z = |z - z^\prime|$, in the central region $z=[-36,36] \, \mu\text{m}$. 
It is evident that enhanced global number squeezing slows down the decay of PCF over large spatial separations $\overline z$. 

For prethermalised states \cite{Gring_2012}, it was found that the local number squeezing is directly linked to the effective temperature $T^-$ of the implemented sine-Gordon model according to the relation $T^- \propto \xi_{N}^2$ (see \cite{erne2018far} and Appendix); thus for decoupled DWs enhanced squeezing leads to a larger phase coherence length $\lambda_T \propto 1/T^-$.

In multimode systems, different spatial modes can in principle feature different levels of squeezing; this would result in mode-dependent effective temperatures as for example in a generalized Gibbs ensemble \cite{langen2015experimental}.
To examine the spatial dependence of squeezing in our experiment, we track the time evolution of the fluctuations $\Delta \theta(z, -z)$ of the relative phase field between two symmetric points. 
We extract linear rates $\partial_t \Delta \theta (z, -z)$ and show an exemplary at two distances $z = 8\mu$m and $24\mu$m in Fig.\,\ref{fig:pcf}\textbf{c}. 
We observe a spatial dependence of dephasing rates $\partial_t \Delta \theta (z, -z)$ (see Fig.\,\ref{fig:pcf}\textbf{d}). 
As expected, the experimental set with better global number squeezing yields an overall lower dephasing rate $ \partial_t \Delta \theta (z, -z)$ and additionally we observe slower dephasing at small distances.\par  

\section*{Conclusion and outlook}
We have observed oscillatory dynamics of quantum fluctuations on conjugate variables of a multimode BJJ and based on this observation, developed a more efficient approach for achieving enhanced spin-squeezed states. 
We envision a more efficient preparation of spin-squeezed states with the help of optimal control algorithms optimizing the classical external dynamics after rapid two-step sequences.
In consideration of our $1$D multimode system, we have demonstrated the influence of number squeezing on prohibiting local dephasing in decoupled DWs. 
In the future, the ability to track the squeezing dynamics provides a new way for optimizing the preparation of strongly correlated Sine-Gordon field simulators with lower effective temperatures; by experimentally approaching a regime that is dominated by quantum fluctuations, measurements of the entanglement entropy in quantum fields will become possible \cite{Tajik_mutualinfo}.\par

\noindent\textbf{Acknowledgements}\\
We thank Sebastian Erne, Camille Lévêque, and Igor Mazets for discussions and Philipp Kunkel for comments on the manuscript.
This work is supported by the DFG/FWF CRC 1225 'ISOQUANT', (Austrian Science Fund (FWF) P~36236) and the QUANTERA project MENTA (FWF: I-6006). M.P. has received funding from the European Union’s Horizon 2020 research and innovation program under the Marie Skłodowska-Curie grant agreement No 101032523.\\
\noindent\textbf{Author contributions}, 
T.Z. and M.P. took the experimental data with the help of Y.K., M.M. and F.B. T.Z. analysed the data with the help of M.P. T.Z., Y.K., J.S. and M.P. discussed the experimental findings and wrote the manuscript with the input from all authors. J.S. and M.P. supervised the work.

\noindent\textbf{Author information}\\
Correspondence and requests should be addressed to T.Z. (tiantian.zhang@tuwien.ac.at) or M.P. (maximilian.pruefer@tuwien.ac.at).

\noindent\textbf{Data availability}\\
Source data and all other data that support the plots within this paper and other findings of this study are available from the corresponding author upon reasonable request.

\noindent\textbf{Competing financial interests}\\
The authors declare no competing financial interests.
\par

\bibliography{references.bib}

\newpage

\setcounter{figure}{0}
\renewcommand{\figurename}{\textbf{Ext.\,Data Fig.}}

~ 
\newpage

\section*{Appendix}\label{sec:Appendix}

\subsection*{Number squeezing factor {estimation}}\label{sec:numsqfacs}
\begin{figure}[h!]
    \centering
    \includegraphics[width = 0.45\textwidth]{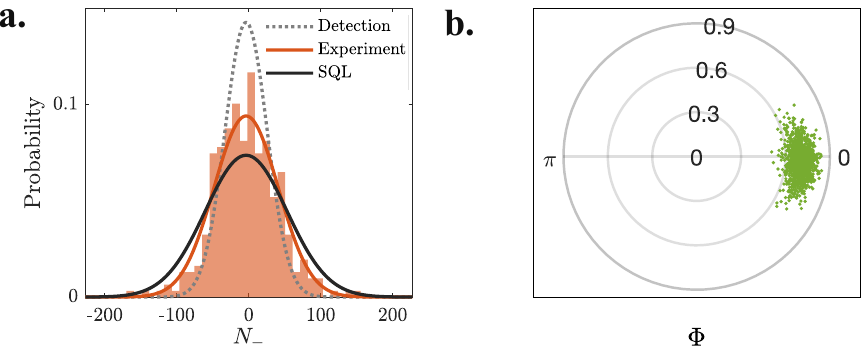}
    \caption{\textbf{Statistical evaluation of distribution} \textbf{a.} Histogram of $N_-$ together with fitted normal distribution (orange). As a reference, we show a binomial distribution (black; indicating the standard quantum limit) and a normal distribution with the width corresponding to the detection noise level (grey). 
    \textbf{b.} Polar distribution of $\Phi$ with fringe visibility $C$ as radii. 
    Errorbars represent one standard error of mean (s.\,e.\,m.).}
    \label{fig:histos}
\end{figure}
We sum the signal from both cloud to obtain the global number of counts $S_L$ and $S_R$ and calculate the global number squeezing factor by
\begin{equation}
    \xi_N^2 = \frac{\Delta^2 S_- - 2S - 2\Delta^2 b}{\overline{p}S}\,.
\end{equation}
Here $\Delta^2 S_- = \Delta^2 (S_L-S_R)$ is the variance on the relative photon signal, $\overline{p}$ is the experimentally calibrated average number of photons collected per atom and $\Delta^2 b$ is the variance of the background noise of an atom-free-region on the EMCCD chip. The noise of $2S = 2(S_L+S_R)$ originates from the electron multiplication process of the EMCCD chip. $2S+\Delta^2b$ is the combined detection noise which is indicated as grey dotted line (Gaussian with corresponding width) in the histogram of $N_-$ in {Ext.\,Data Fig.\,\ref{fig:histos}\textbf{a}}.\par

\subsection*{Extraction of relative phase}\label{sec:phsqfacs}
We fit the interference fringe slice-wise with $\rho(x) \approx g(x) [1 + C \cos(k_0(x-x_0) +\phi)]$ to extract the local relative phase $\phi$ and the fringe visibility $C$, indicating the single particle coherence.
To minimize the readout error originating from the locally fluctuating relative phases, we extract the global relative phase $\Phi = \text{arg} \left( \overline{\exp{(i\phi(z))}} \right)$, based on independently fitted local relative phase $\phi (z)$ over region $z = [-3, 3]\delta z$, {where $\delta z = 4 \, \mu \text{m}$ is the pixel size in the object space}. {The evaluated phase distribution is shown in Ext.\,Data Fig.\,\ref{fig:histos}\textbf{b}.}\par

\subsection*{Characterization of imaging resolution}
\begin{figure}[h!]
    \centering
    \includegraphics[width = 0.5\textwidth]{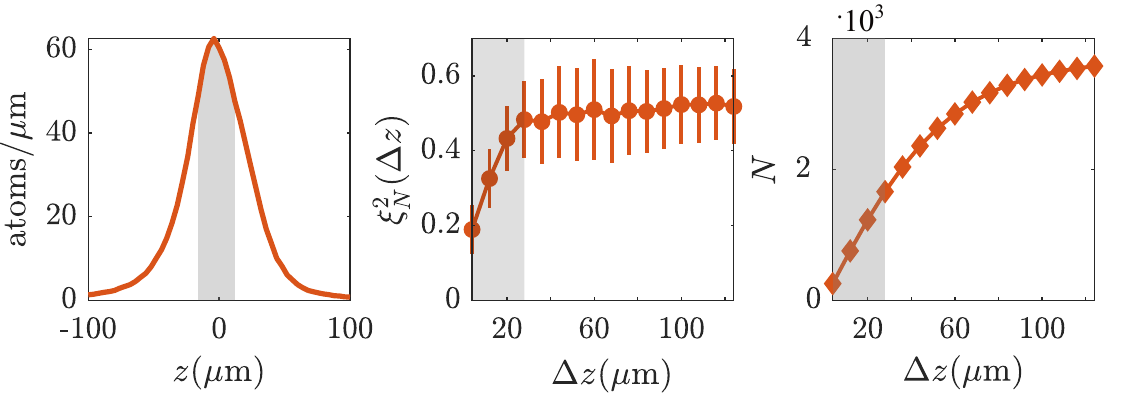}
    \caption{\textbf{Imaging influence on spatially resolved number squeezing detection}. Left panel: Longitudinal profile of BEC.  Middle panel: Influence of finite integration length $\Delta z$ on evaluation of number squeezing factor $\xi_N^2$. $\Delta z$ is evaluated symmetrically around the peak density along the condensate. Right panel: Atom number $N$ within integration region $\Delta z$. The shaded region indicates the lower bound of $\Delta z$ for reliable estimation of $\xi_N^2$.}
    \label{fig:localnumsqueezing}
\end{figure}

The longitudinal extension of the condensates is $L = 60 - 120\, \mu \text{m}$ depending on the total atom number. Due to the random walk of atoms in the imaging light and diffusion of emitted photons, the effective imaging resolution is larger than the imaging resolution $\delta z$. The effective imaging resolution is critical for the evaluation of the local number squeezing factor. We show in Ext.\,Data Fig.\,\ref{fig:localnumsqueezing} that the calculated $\xi_N^2$ reaches a steady value with integration regions above $28\, \mu \text{m}$. This length signifies the effective resolution which is still a few times smaller than the condensate length.\par

\subsection*{Mean-field Josephson oscillation}

\begin{figure}[h!]
    \centering
    \includegraphics[width = 0.35\textwidth]{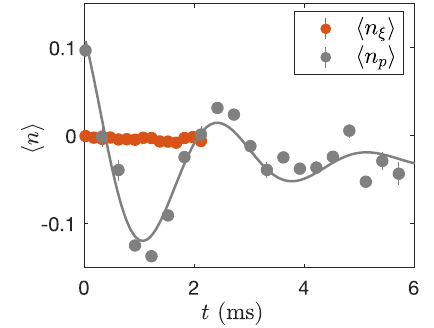}
    \caption{Evolution of expectation value of relative imbalance $\langle n \rangle = \langle N_- / N \rangle$ in coupled DW. 
    Grey markers, $\langle n_p \rangle$, show Josephson oscillation with an imprinted nonzero initial imbalance. Orange markers, $\langle n_\xi \rangle $, indicate that the expectation values are at equilibrium after symmetrical splitting.}
    \label{fig:JosephsonExample}
\end{figure}

We show in Ext.\,Data Fig.\,\ref{fig:JosephsonExample} an example of Josephson oscillation of relative imbalance $\langle n_p\rangle$ by imprinting an initial nonzero imbalance, which is used for extraction of the plasma frequency $f_p$ (shown in Fig.\,\ref{fig:drivefreq}\textbf{c}).\par
For comparison, we also display the typical evolution of the imbalance, $\langle n_\xi \rangle \approx 0$, in the case of symmetric splitting {investigated} in this work.\par 

\subsection*{Fit and errorbars} \label{sec:fit_err}
Squeezing oscillations are fitted with the function
$\xi^2(t) = a\cdot \sin(2\pi f_\xi \cdot t + p_0) + \overline {\xi^2}$ with $a, f_\xi, p_0$ and $\overline{\xi}$ as its fit parameters. The shaded region of the fitted model signifies $68\%$ simultaneous bound prediction confidence intervals of the fit function. Error bars on experimentally measured squeezing factors are estimated standard error using a jackknife resampling. 

\subsection*{{Semiclassical} simulation on squeezing oscillations} \label{sec:simulation}

\begin{figure}
    \centering
    \includegraphics[width = 0.5 \textwidth]{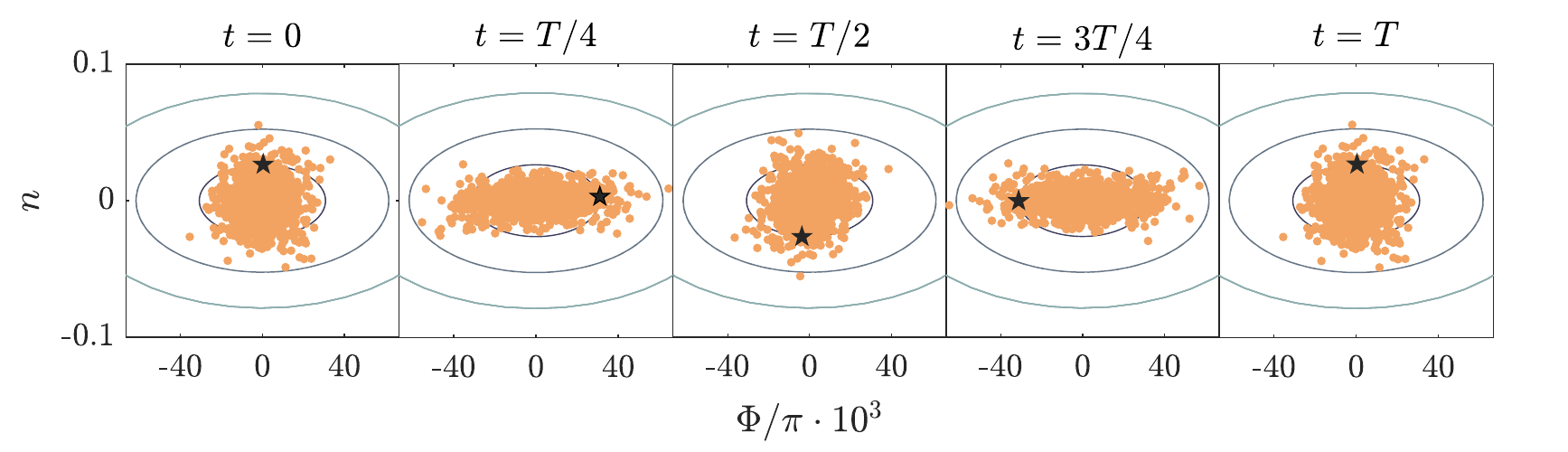}
    \caption{\textbf{Propagation of imprinted initial fluctuations in the two-mode BH model}. The star marker signifies the evolution of a single realisation, representing the mean field value. A $\pi$ rotation of a single realisation corresponds to a $2\pi$ rotation of the phase space fluctuations. Here $T$ is one period of Josephson oscillation.}
    \label{fig:simphasespace}
\end{figure}

To visualize the squeezing oscillation in the coupled DW (see Fig.\,\ref{fig:conjosci}), we set up a semi-classical simulation. In Ext.\,Data. Fig. \ref{fig:simphasespace}, we plot the equipotential lines in the classical phase space given by the two-mode Bose-Hubbard model
\begin{equation}\label{exteq:bose-hubbard}  
 \mathcal{H} = \frac{2J}{\hbar} \left[ \frac{{\Lambda}}{2}  n^2 - \sqrt{1- n^2} \cos {\Phi}\right],
\end{equation} 
{where $\Lambda = UN/2J$ signifies the interplay between the inter-atomic interaction energy and tunnel coupling energy}.
With suitable parameter values for DW $\mathbf{\mathcal{A}} = 0.5$: single particle tunnel coupling energy $J = 41\,\text{Hz}$, interaction energy $U = 0.33 \,\text{Hz}$ and total atom number $N = 3500$. Here $J$ is estimated using the energy difference between the two lowest single-particle eigenstates in the simulated DW potential, $J = (E_1-E_0)/2$, and experimentally measured atom number $N$.\par 

\begin{figure}[h!]
    \centering
    \includegraphics[width = 0.45\textwidth]{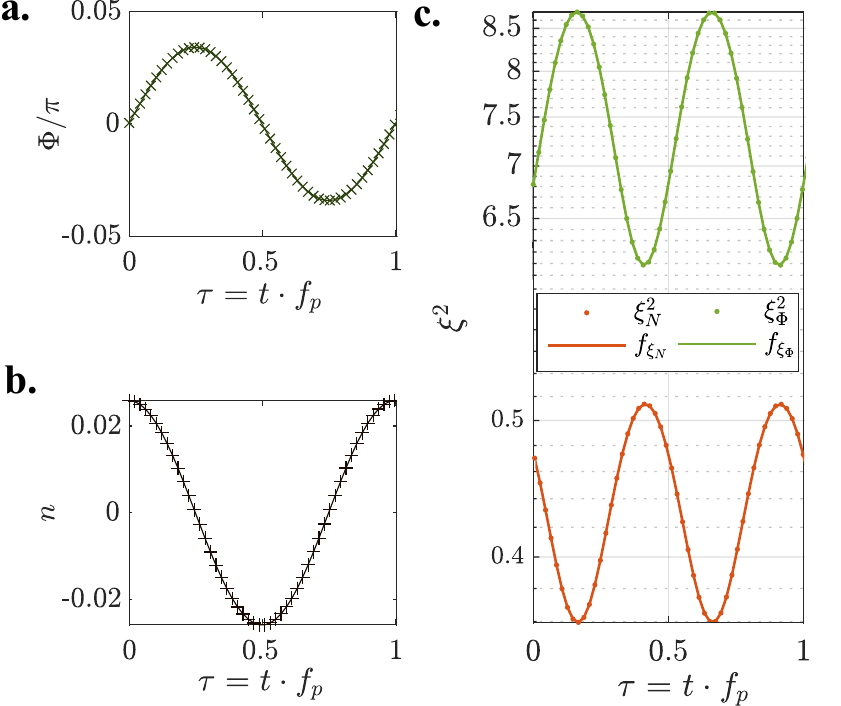}
    \caption{\textbf{Squeezing oscillation frequency} Evolution of a single realisation in Ext.\,Data Fig.\,\ref{fig:simphasespace} in the $\Phi$ (in \textbf{a}) and $N_-$ quadrature (in \textbf{b}). Evolution of quantum fluctuation in the conjugate quadratures (in \textbf{c}) and its fitted frequency corresponding to twice the plasma frequency $f_p$.}
    \label{fig:simsqfac}
\end{figure}
Around the stable fixed point and given small fluctuations in the lower energy states, BH Hamiltonian can be further linearized and expressed in harmonic approximation as 
\begin{equation} \label{eq:bhhc}
     H_{hc} = \frac{h f_p}{2} \left( \frac{ \Phi^2}{2\Delta_{GS}^2 \Phi} + \frac{ N_-^2}{2\Delta_{GS}^2 N_-}\right), 
\end{equation}
where $\Delta_{GS}^2 \Phi = \sqrt{1+\Lambda}/N$ and $\Delta_{GS}^2 N_- = N/\sqrt{1+\Lambda}$ are ground state fluctuations. With these parameters, we can estimate the expected ground state squeezing factors in Eq.\,\eqref{eq:bhhc} to be
\begin{equation}\label{eq:sgsqfac}
 \xi_{N,0}^2 = 1/\sqrt{1+\Lambda}, \hspace{5mm}\xi_{\Phi,0}^2 = \sqrt{1+\Lambda},   
\end{equation}
 {shown as shaded bands in Fig.\,\ref{fig:expsetup}\textbf{b}}.\par
 
We sample $1000$ realisations from two normal distributions (one for each observable)) with variances larger than the ground state fluctuations $\Delta^2_{GS}$ (deduced from Eq.\,\eqref{eq:bhhc}) and propagate them with equations of motion deduced from Eq.\, \ref{exteq:bose-hubbard} in the classical limit. 
We show in Ext.\,Data\,Fig.\,\ref{fig:simphasespace} the simulation result of quantum state propagation in a time span of $T = 1/f_p$, where $T$ corresponds to a period of Josephson oscillation of the expectation value of the obsevables (star marker). {To make this more explicit, we plot the evolution of $N_-$ and $\Phi$ of a single realisation and the projected fluctuations as squeezing factor $\xi^2$ in each quadrature in Ext.\,Data Fig.\,\ref{fig:simsqfac}.} As one can see, the projection noise in each observable oscillates at twice the frequency as the expectation values, namely $f_\xi = 2 f_p$ (see also \cite{dunningham2001relative}).\par 

\subsection*{Impact of number squeezing on global phase diffusion}
 In decoupled trap ($J\approx 0$), phase diffusion after symmetric splitting \cite{javanainen1997phase,leggett1998comment} can be expressed as
\begin{equation}\label{eq:phasediffusion}
     \Delta^2 \Phi (t) = \Delta^2\Phi_0 + R^2 t^2,  
\end{equation} 
where $R = \frac{\xi_N \sqrt{N}}{\hbar} \frac{\partial\mu}{\partial \mathcal{N}}\Big|_{\mathcal{N} = N/2}$ is the phase diffusion rate and $\Delta^2 \Phi_0$ is the initial variance of $\Phi$ right after splitting and $\mu(\mathcal{N})$ is the chemical potential of BEC with atom number $\mathcal{N}$. 
Eq.\,\eqref{eq:phasediffusion} implies slower phase diffusion with stronger number squeezing.\par 

\begin{figure}
    \centering
    \includegraphics[width = 0.5\textwidth]{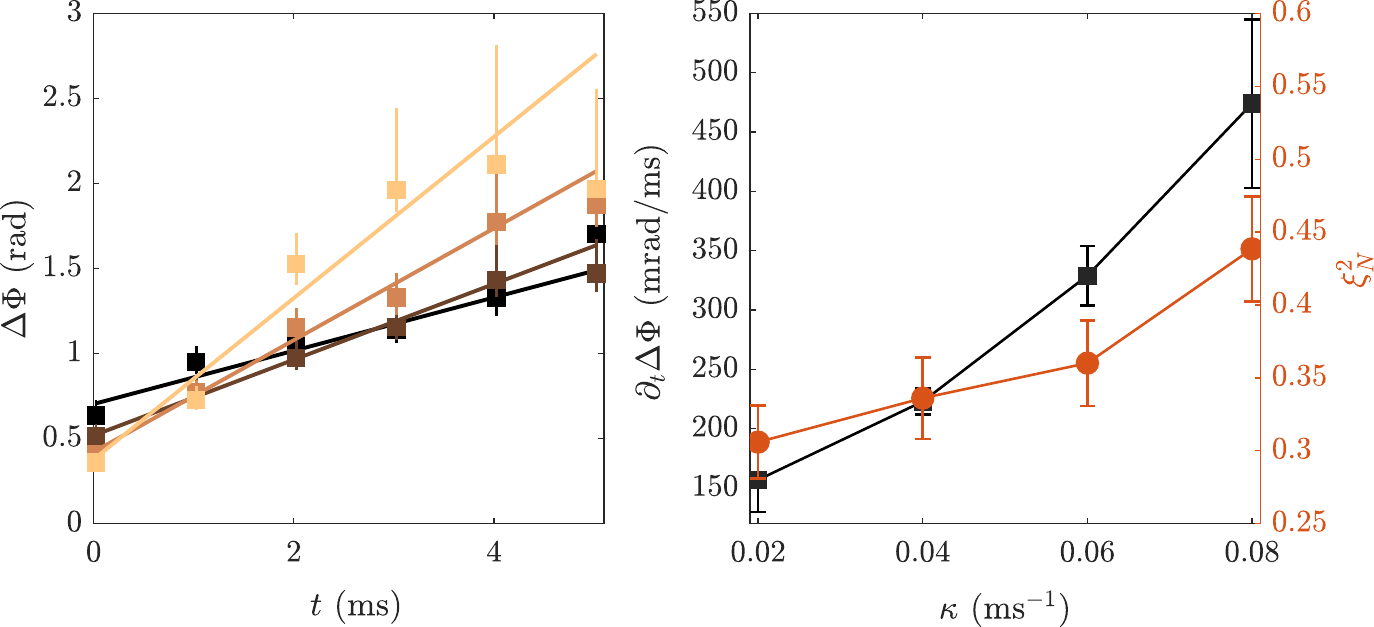}
    \caption{\textbf{Global number squeezing suppresses global phase diffusion in decoupled trap}. Left panel: phase diffusion in $\mathbf{\mathcal{A}} = 0.6$ trap for different splitting speeds $\kappa$. Right panel: The linearly fitted phase diffusion rates $\partial_t \Delta \Phi$ agree qualitatively with the measured $\xi_N$ with different $\kappa$.}
    \label{fig:split_speed}
\end{figure}

We investigate experimentally the influence of number squeezing on global phase diffusion rate by splitting into effectively decoupled DW, $\mathbf{\mathcal{A}} = 0.6$. For different split speed $\kappa  =\delta \mathbf{\mathcal{A}} /\delta t $ we measure the phase spread $\Delta \Phi (t)$ and deduce the phase diffusion rate from a linear fit.
The extracted rates $\partial_t \Delta \Phi$ match the trend of the measured $\xi_N$. (see Ext.\,Data.\,Fig.\, \ref{fig:split_speed}).\par

\subsection*{Impact of number squeezing on spatial dephasing} \label{sec:locphasediff}
For resolving local dynamics between two split 1D condensates, we introduce the local observables along the condensates: the local relative phase $\phi(z) = \phi_L(z) - \phi_R(z)$ and the relative density $\rho_-(z) = \rho_L(z) - \rho_R(z)$. 
The local and global observables fulfill the relation $\Phi = \text{arg} \left( \overline{\exp{(i\phi(z))}} \right)$, $N_- = \sum_z \rho_-(z)$.\par

\begin{figure}
    \centering
    \includegraphics[width = 0.5\textwidth]{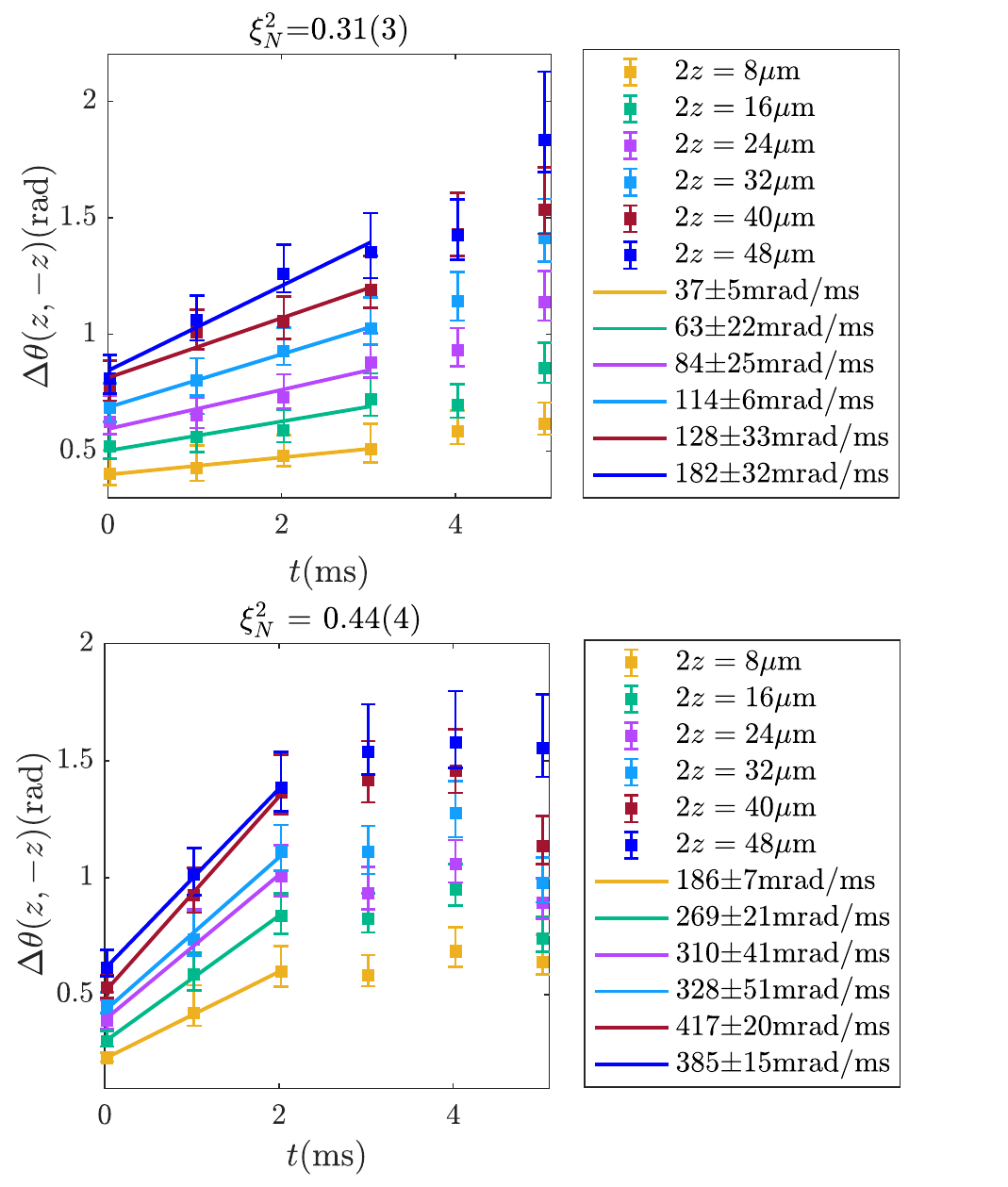}
    \caption{\textbf{Extraction of local dephasing rates in decoupled trap in relation to global number squeezing}. Evolution of spatial relative phase fluctuations $\Delta \theta(z,-z)$ between two symmetrically located points around the longitudinal center of the BEC. Upper panel: $\Delta \theta(z,-z)$ evolution with global number squeezing $\xi_N^2 = 0.31(3)$ and a linear fit on the four first time instances. Lower panel: $\Delta \theta(z,-z)$ evolution with global number squeezing $\xi_N^2 = 0.44(4)$ and a linear fit on the three first time instances.}
    \label{fig:localphasediff}
\end{figure}

In previous works studying the phenomenon of prethermalisation \cite{Gring_2012,kitagawa2011dynamics}, the effective temperature $T^-$ between the two condensates  was connected to the relative density fluctuations $\Delta^2\rho_- =\langle \rho_-^2\rangle$ by 
\begin{equation}
    T^- = \frac{g \Delta^2 \rho_-}{2} = \frac{g\rho_0\xi_\rho^2}{2}\,,
\end{equation}
where $\rho_0$ is the peak atomic density in each condensate after splitting and $\xi_\rho^2$ is the local number squeezing factor.\par

\end{document}